\documentclass[amsmath,amssymb]{revtex4}
\usepackage{mathrsfs}
\usepackage[dvips]{graphicx}
\begin{document}
\author{M.Dowling, A.Pak}
\affiliation{Department of Physics, University of Alberta, Edmonton AB T6G 2J1, Canada}
\title{Massive $c$-quark contributions to semileptonic $b$-quark decay near threshold}
\date{\today}
\begin{abstract}
\begin{center} \large{Abstract} \end{center}
The decay width of $b\to c\overline{c}c \ell \overline{\nu}_{\ell}$ has been computed as an expansion in limit $(1- 3{m_c\over m_b}) \ll 1$ with zero lepton invariant mass.
Considering $\mathcal{O}(\alpha_s^2)$ corrections to $b\to c \ell \overline{\nu}_{\ell}$ with zero invariant lepton mass, inclusion of our result is shown to rectify the disagreement between the expansions around $(1- {m_c \over m_b}) \ll 1$ and ${m_c \over m_b} \ll 1$ at smaller values.
\end{abstract}
\maketitle
\newpage
\section{Introduction}
Recent experimental and theoretical progress in measuring the semileptonic $b$-quark decay width $\Gamma(b \to X \ell \overline{\nu}_{\ell})$ has allowed us to measure fundamental parameters such as $|V_{cb}|$ with good accuracy.
The theoretical precision of this decay relies in part on good knowledge of perturbative QCD corrections which include two relevant scales and are thus challenging to calculate.
In papers \cite{PhysRevLett.78.3630,pak:114009} order $\mathcal{O}(\alpha_{s}^2)$ corrections to this decay have been presented as series in terms of small parameters ${m \over M}$ and $(1- {m\over M})$ respectively, in the kinematic limit of zero lepton pair invariant mass with $m = m_c$ and $M = m_b$.
This present work is targeted to reconcile those two expansions in the region $m < {M \over 3}$ where the expansion in \cite{PhysRevLett.78.3630} was missing the contribution of the process $b \to c\bar{c}c\ell\bar{\nu}_{\ell}$.

In this paper we present the contribution of this decay mode as a series in $\delta = 1 - 3{m \over M}$, and demonstrate that with this correction the results of \cite{PhysRevLett.78.3630} match those of \cite{pak:114009} to within 1\% in almost the whole allowed region.

In the section \ref{sec:decay} we will present relevant details of the calculation and introduce notation used, section \ref{sec:results} will be used to discuss the obtained expansion.

\section{\label{sec:decay}Decay width of $b \rightarrow c \overline{c} c W^*$}

\begin{figure}[!hbt]
\centering
\includegraphics[width=0.4\textwidth]{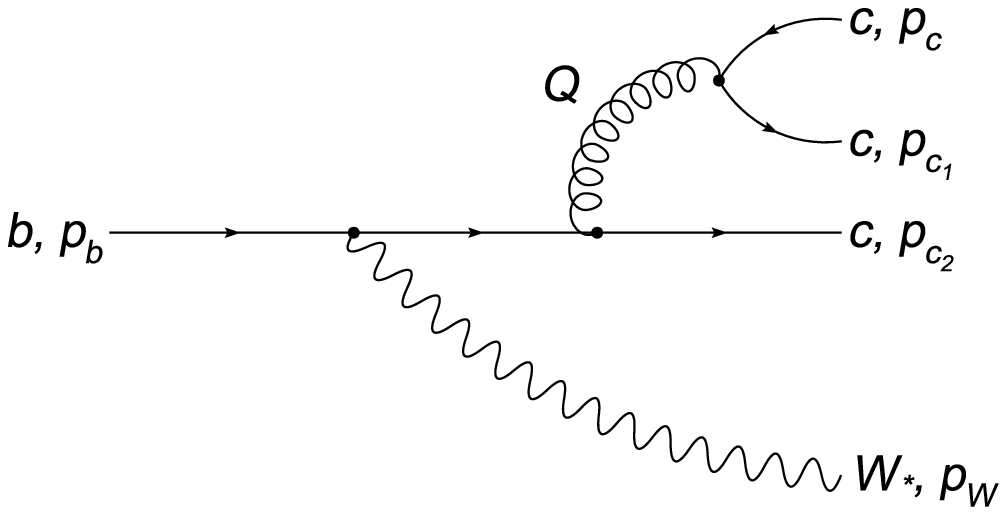}
\includegraphics[width=0.4\textwidth]{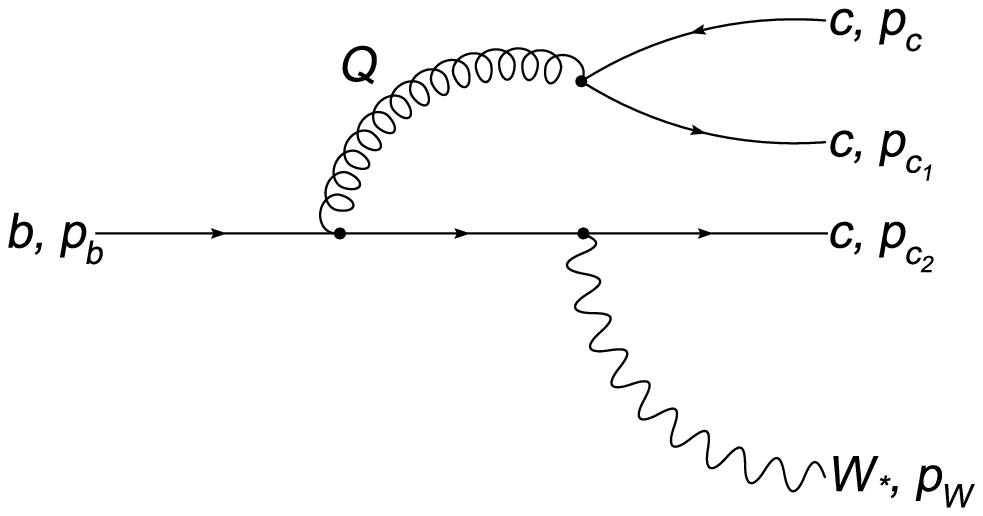}
\caption{\label{feyn}Tree-level Feynman diagrams contributing to $b \rightarrow c \overline{c} c W^*$}
\end{figure}

As shown in \cite{pak:114009}, the decay width $d\Gamma(b \rightarrow c\overline{c}c \ell \overline{\nu}_{\ell})/dq^2$ in the limit of zero lepton invariant mass, $q^2 \rightarrow 0$, is equivalent to $d\Gamma(b \rightarrow c\overline{c}cW^*)$ with $m(W^*)=0$ up to a constant factor.
The two Feynman diagrams that contribute to this process along with notations for particle momenta are shown in Fig. \ref{feyn}.
We define its width as
\begin{eqnarray}
\Gamma(b \rightarrow c \overline{c}cW^*) &=& \int {(2\pi^4) \over 2M} \langle |M_T|^2 \rangle dR_4(p_b;p_{c_2},p_W,p_{c_1},p_{\overline{c}}), \\
dR_n(p_0;p_1,\ldots,p_n) &=& \delta^4(p_0 - \sum_{i=1}^n p_i) \prod_{i=1}^n {d^3p_i \over (2\pi)^3 2 E_i}.
\end{eqnarray}

Introducing $Q = p_{c_1} + p_{\overline{c}}$ and $U = Q + p_W$, we can factorize the phase space element as follows:

\begin{eqnarray} 
  dR_4(p_b;p_{c_2},p_W,p_{c_1},p_{\overline{c}}) &=& 
  dR_2(p_b;U,p_{c_2}) \\ \nonumber
  &\times& dU^2(2\pi)^3 dR_2(U;Q,p_W)dQ^2(2\pi)^3
 dR_2(Q;p_{c_1},p_{\overline{c}}) 
\end{eqnarray}

To integrate the matrix element $\langle |M_T|^2 \rangle$, we introduce the small parameter $\delta = 1 - 3\frac{m}{M}$, with $\delta=0$ corresponding to the threshold of the decay and $\delta = 1$ limiting the radius of convergence of the expansion due to IR divergences.

Comparing the relative scales of different combinations of momenta we may expand the denominators as follows:

\begin{eqnarray} 
\frac{1}{(p_{b} - p_{W})^2 - m^2} &=& {1 \over 8m^2} - {2(p_{c_1} \cdot p_{\overline{c}} + Q \cdot p_{c_2} - 3m^2) \over (8m^2)^2}  \\ \nonumber
  &+&  {4(p_{c_1} \cdot p_{\overline{c}} + Q \cdot p_{c_2} - 3m^2)^2\over (8m^2)^3} - \ldots, \\
 {1\over (p_W + p_{c_2})^2 - M^2} &=& -{1 \over 4m(M - m)} - {2(p_{c_1} \cdot p_{\overline{c}} - p_b \cdot Q -m^2 +2Mm) \over (4m(M-m))^2} \\ \nonumber
  &-& {4(p_{c_1} \cdot p_{\overline{c}} - p_b \cdot Q -m^2 +2Mm)^2\over (4m(M-m))^3} - \ldots
\end{eqnarray}

Since this expansion is done at threshold the decay products from each two particle decay are produced almost at rest, which allows us to expand scalar products e.g. $p_{c_1} \cdot p_{\overline{c}} \to m^2$, $p_b \cdot Q \to 2Mm$, and $Q \cdot p_{c_2} \to 2m^2$.

Now the denominators contain only constants so the phase space integrals are easier to deal with. 
After taking traces and expanding the propagators the $R_2$ integrals are all in the form
\begin{equation}
\int dR_2(P;q_1,q_2)(P\cdot q_1)^a(P\cdot q_2)^b.
\end{equation}
We treat those integrals as explained in \cite{PhysRevD.56.7216}(Sect.III.A).
The decay width now becomes:

\begin{equation} \Gamma(b\to c\overline{c}cW^*) = \int\int dU^2 dQ^2  {2\pi^4 \over 2M} X_1(Q^2) X_2(Q^2,U^2) X_3(Q^2,U^2) \langle |M_T|^2 \rangle(U^2,Q^2)
\end{equation}

with

\begin{equation}
\frac{X_1(Q^2)}{(2\pi^6)}=\int dR_2(Q;p_{c_1},p_{\overline{c}}), ~~~
\frac{X_2(U^2,Q^2)}{(2\pi^6)}=\int dR_2(U;Q,p_{W}), ~~~
\frac{X_3(Q^2,U^2)}{(2\pi^6)}=\int dR_2(p_0;U,p_{c_2}).
\end{equation}

The remaining integrals over $U^2$ and $Q^2$ have limits $4m^2 < U^2 < (M-m)^2$ and $ 4m^2 < Q^2 < U^2$. 
We introduce $x_1,x_2 \in [0,1]$ such that $Q^2 = 4m^2(1 + \omega x_1)$ and $U^2 = 4m^2(1 + \omega x_1x_2)$, with
$\omega = {3\delta(4-\delta) \over 4(1-\delta)^2}$, and $X_{1,2,3}$ become:
\begin{equation}
X_1 = {\pi \over 2} \sqrt{w} \sqrt{x_1 x_2} \sqrt{{1 \over 1 + \omega x_1 x_2}} , ~~~
X_2 = {\pi \over 2} {\omega x_1(x_2 - 1) \over 1 + \omega x_1 x_2}, ~~~
X_3 = {2\pi m \over \sqrt{3}M} \sqrt{w} \sqrt{1-x_1}\sqrt{1 - {\delta^2 \over 4} - {x_1 \delta \over 4}(4 - \delta)}
\end{equation}

Since $\omega \sim \delta$ and $0 \leq x_1,x_2 \leq 1$ the term $\omega x_1 x_2$ is small and can be used as an expansion parameter in the first two volumes.
The last square root in the third volume can also be expanded because of the small terms that depend on $\delta$.
These expansions simplify the integrals over $x_1$ and $x_2$ to well known forms so that they can easily be carried out.

\section{\label{sec:results}Results}

Here we present several terms in the expansion of the result in $\delta$.

\begin{eqnarray} 
\Gamma(b\rightarrow c \overline{c} c W^*) &=& \Gamma_0 \alpha_s^2 C_F T_R Y_c, \\
\label{result}            
 Y_c &=& {\sqrt3 \over 4 \pi}
              \delta^6 \left(
              {4 \over 5} +
              {83 \over 70}\delta +
              {7 \over 80}\delta^2 +
	      {11 \over 224}\delta^3 +
	      {753 \over 1120}\delta^4 \ldots \right),
\end{eqnarray}

where $C_F={4\over 3}$ and $T_R={1\over 2}$ are color factors in $SU(3)$, $\Gamma_0 = {G_F|V_{cb}|^2M^3 \over 8\sqrt2 \pi}$, and
$G_F = {g_w^2 \sqrt2 \over 8 M_w^2}$.

\begin{figure}[!hbt]
\centering
\includegraphics{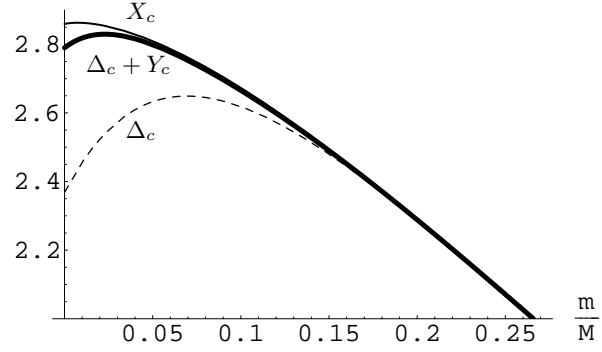}
\put(-185,80){$\Delta_c$}
\put(-200,105){$\Delta_c + Y_c$}
\put(-185,125){$X_c$}
\caption{\label{plot}Comparison of results from previous papers. The dashed line shows the expansion about ${m \over M} = 1$ while the thin line shows the expansion about ${m \over M} = 0$.
The thick line shows how the result from this paper closes the gap between the two.}
\end{figure}

Fig. \ref{plot} shows functions $X_c$ of \cite{pak:114009} and $\Delta_c$ of \cite{PhysRevLett.78.3630} (as given in \cite{pak:114009}) along with $\Delta_c + Y_c$ with our expansion (\ref{result}) taken to $\delta^{14}$ all depending on $\frac{m}{M}$. 
The good agreement of the solid curves is an indication of the consistency of expansions in \cite{PhysRevLett.78.3630} and \cite{pak:114009}.

\section{Conclusion}
In this paper we presented a computation of the decay width $\Gamma(b \rightarrow c\overline{c}cW^*)$ in the limit $\left(1-3{m \over M}\right) \ll 1$.
The inclusion of our result to $\mathcal{O}(\alpha_s^2)$ corrections computed in \cite{PhysRevLett.78.3630,pak:114009} closes the gap between the two calculations with a maximum discrepancy of 2.5\% due to a limited number of terms calculated in the expansions.
This, once again, shows that expansion methods are a powerful tool in dealing with multi-loop and multi-scale problems.

Acknowledgments:
We would like to thank Andrzej Czarnecki for his support throughout the project.
For a large part of the calculations the symbolic algebra program FORM \cite{vermaseren-2000} was used.
This research was supported by Science and Engineering Research Canada.

\bibliography{3charm}

\end{document}